\documentstyle[12pt,epsf]{article}
\begin{document}
\begin{titlepage}
\vspace{5mm}
\begin{center}
{\large \bf Spontaneous $CP$ Violation and Higgs Masses \\ 
in the Next-to-Minimal Supersymmetric Model } \\ %
\vspace{15mm}
Naoyuki HABA$^{a,}$\footnote{E-mail:\ haba@eken.phys.nagoya-u.ac.jp}, 
Masahisa MATSUDA$^{b,}$\footnote{E-mail:\ mmatsuda@auecc.aichi-edu.ac.jp} 
and
Morimitsu TANIMOTO$^{c,}$\footnote{E-mail:\ tanimoto@edserv.ed.ehime-u.ac.jp} \\
{\it
${}^a$Department of Physics, Nagoya University \\ 
Nagoya, JAPAN 464-01 \\
${}^b$Department of Physics and Astronomy \\ 
Aichi University of Education \\
Kariya, Aichi, JAPAN 448 \\
${}^c$Science Education Laboratory, Ehime University, \\ 
Matsuyama, JAPAN 790 \\
}
\end{center}
\vspace{10mm}
\begin{abstract}
We study the possibility of spontaneous $CP$ violation in 
the next-to-minimal supersymmetric standard model (NMSSM). 
It is shown that the spontaneous $CP$ violation is induced by 
the radiative effects of top, stop, bottom and sbottom superfields. 
The available regions of parameters, which are obtained by imposing 
the constraints from experiments, are rather narrow.
We also obtain strong constraints for light Higgs masses such as $m_H 
\le 36$GeV numerically.
Sum of masses of
two light neutral Higgs should set around 93GeV and 
charged
Higgs boson has a rather higher mass larger than 700GeV.
\end{abstract}
\end{titlepage}
\section{Introduction}
The physics of $CP$ violation has attracted much attention 
in the light that the $B$-factory
will go on line in the near future at KEK and SLAC. 
The central subject of the $B$-factory is the test of the standard model(SM),
in which the origin of $CP$ violation is reduced to the phase in the 
Kobayashi-Maskawa matrix\cite{KM}. However, there has been a general 
interest in considering other approaches to $CP$ violation since many 
alternate sources exist. 
The attractive extension of the
standard Higgs sector is the two Higgs doublet model(THDM)\cite{THDM}, 
yielding both charged and neutral Higgs bosons
as physical states. The THDM with the soft breaking term of the discrete 
symmetry demonstrates explicit or spontaneous $CP$ violation 
\cite{LEE}\cite{BRANCO}\cite{WEINBERG}.
On the other hand, the recent measurements of  gauge couplings at 
$M_Z$ scale suggest the minimal supersymmetric extension of the standard model(MSSM) is
a good candidate
beyond the standard model in the standpoint of the gauge unification\cite{AMALDI}.

It is well known that  $CP$ symmetry could 
be violated explicitly or spontaneously in the THDM without supersymmetry\cite{HAYASHI}.
Though the MSSM contains two Higgs doublets $H_1^T=(H_1^{0},H_1^-)$ and 
$H_2^T=(H_2^+,H_2^0)$,
which give masses to down-quarks and up-quarks, respectively, there is 
no degree of freedom for $CP$ violation at tree level Higgs potential. 
Spontaneous symmetry breaking of $SU(2)_L$ by taking non-zero real 
vacuum expectation values(VEV)
gives rise to two $CP$-even neutral Higgs scalars, a $CP$-odd neutral 
pseudo-scalar boson, and two charged Higgs bosons.
One of two $CP$-even bosons is the lightest of all Higgs bosons in the 
MSSM and its tree level
mass is less than that of $Z^0$.
However, large radiative corrections proportional to $(g^2m_t^4/M_W^2)$ 
increase the lightest mass of the neutral Higgs bosons of the MSSM  than $M_Z$\cite{OKADA}.
Within a framework of the MSSM it is also possible to violate $CP$ symmetry 
spontaneously by radiative effects of heavy quarks 
with relatively non-zero complex VEVs for $H_1^0$ and $H_2^0$\cite{MAEKAWA}. 
Phenomenologically this model requires the lightest mass 
of the neutral Higgs boson to be a few GeV 
as a  result of Geogi-Pais theorem\cite{GP}.
So this interesting scenario to violate $CP$ symmetry spontaneously in the 
MSSM is unfortunately inconsistent with the experiment which suggests that 
the lightest pseudoscalar Higgs mass is larger than 22GeV\cite{PDG}. 
To avoid this difficulty, the simple extension of the MSSM 
has been considered to obtain explicit\cite{MATSUDA} or 
spontaneous\cite{BABU} $CP$ violation in the Higgs sector. 
The extension is that a singlet superfield under $G_{st}=SU(3)_C\times SU(2)_L\times 
U(1)_Y$ is added to the MSSM. This model is usually called as next-to-minimal 
supersymmetric standard model(NMSSM)
\cite{FAYET}\cite{RG}.

The NMSSM is introduced to solve so called $\mu$-problem. 
The superpotential needs the term like $\mu H_1H_2$ to give 
the non-zero VEVs for  both Higgs doublets in the MSSM, where
$\mu$ should be $O(M_W)$. However, the MSSM does not explain 
why $\mu$ should be so small. In the NMSSM we can introduce 
$\lambda NH_1H_2$-term in the superpotential, where $N$ is a 
singlet superfield under $G_{st}$ and $\lambda$ is Higgs 
coupling with $O(\lambda)\simeq 1$. If $N$ develops a non-zero 
VEV $\langle N \rangle\equiv x$, the $\mu$-term is generated as 
$\mu=\lambda x\simeq O(M_W)$. Such a singlet field appears in 
grand unified supersymmetric models\cite{NILLES} and in 
massless sectors of superstring models\cite{COHEN} as well as 
in superstring models based on $E_6$\cite{HABA} and $SU(5) \times U(1)$
gauge groups\cite{ELLIS}.
The minimal extension of the MSSM with an additional singlet 
superfield is an attractive alternative and these models are 
analyzed by many authors with no spontaneous $CP$ 
violation\cite{RADIATIVE}\cite{PANDITA}. 
\par
In the NMSSM, candidates to have non-zero VEVs are $H_1^0,H_2^0$ 
and $N$ and it is likely to develop relatively complex VEVs to 
violate $CP$ symmetry spontaneously.
Furthermore, in order to obtain relatively large mass of neutral Higgs, 
$x$ should be 
rather large compared to $v=\sqrt{v_1^2+v_2^2}=174$GeV,
where $v_1$ and $v_2$ are the VEVs of $H_1^0$ and 
$H_2^0$, respectively.
\par
Recently Babu and Barr have shown that there exists the solution to lift 
the lightest mass of Higgs boson to the consistent region with the present experimental 
lower bound of its mass\cite{BABU}.
In this analysis they pointed out that the spontaneous $CP$ violation 
occurred by the radiative effect of 
stop and top loop in the NMSSM
 for the parameter $\tan\beta=v_2/v_1\simeq 1$.
However, they used the  simplified squark mass matrix as $m_{{\tilde t}_L}=m_{{\tilde t}_R}$ and 
neglected the sbottom and bottom contributions for one-loop correction. 
Furthermore there is one problem that the charged Higgs mass would  be around 100GeV, which might be 
excluded in the minimal supergravity model\cite{GOTO} with the experiment
$b \rightarrow s+\gamma$ \cite{CLEO}.
The charged Higgs mass should be larger than 160GeV in 
this model for small $\tan\beta$, while it's limit is 250GeV in 
the THDM\cite{MATSUDAetal}.
\par
In this paper we introduce the full radiative corrections from
top, stop, bottom and sbottom contribution in the NMSSM 
which derive the different 
results from Ref.\cite{BABU}. 
We also determine the available parameter regions 
in the NMSSM with spontaneous $CP$-violation 
by imposing  precise experimental constraints for the lower limit of neutral Higgs mass
>from $Z \rightarrow h_1+h_2$ and $Z \rightarrow h_1+l^+l^-$ decay processes.
In particular, it is found that the lightest Higgs boson, whose main 
component is pseudoscalar, has a mass with about 36GeV maximally 
and the sum of the masses of two lightest  
Higgs particles is around 93GeV.
So these particles are expected to be observed at LEP2 in the 
near future if the origin of the $CP$ violation in the Higgs 
sector is reduced to the NMSSM with
nontrivial phases of VEVs of two Higgs scalars($H_1^0, H_2^0$) and 
a singlet  scalar($N$).
The mass of charged Higgs is larger than 700GeV, 
which is consistent 
with the present experimental lower limit for the charged Higgs mass.
\par
Section 2 is devoted to the formulation of the NMSSM. In section 3, 
we discuss the framework of the experimental constraints and the spontaneous $CP$-violation 
scenario in the NMSSM.
Section 4 gives parameters of the NMSSM and the masses of neutral 
and charged Higgs bosons by using
the experimental constraints obtained in section 3. 
In section 5 we gives summary and discussions.

\section{Higgs Potential in the NMSSM and Higgs Masses} 

We study the spontaneous $CP$ violation and the Higgs boson masses 
with radiative corrections of top, stop, bottom and sbottom fields in the NMSSM.
Here the radiative effects of top superfield is essential and bottom 
superfield are significant especially in the case of large $\tan\beta$, so that the 
relevant terms in the superpotential is %
\begin{equation}
\label{W}
W = h_tQH_2T^c + h_bQH_1B^c + \lambda N H_1 H_2 + {k \over 3} N^3, \end{equation}
where
\begin{equation}
H_1=\left(
\begin{array}{c}
H_1^{0} \\
H_1^- \\
\end{array}
\right) , \qquad
H_2=\left(
\begin{array}{c}
H_2^+ \\
H_2^0 \\
\end{array}
\right).\nonumber
\end{equation}
with
$$H_1H_2=H_1^{0}H_2^0-H_1^-H_2^+. $$
The cubic term in $N$ is introduced to avoid a Peccei-Quinn symmetry which would
require the existence of a light pseudo-Goldstone boson when the symmetry is broken
by  non-zero VEVs of Higgs fields. 
The superpotential $W$ is  scale invariant and $Z_3$
invariant  which might interpret the weak scale baryogenesis\cite{WHITE}.
\par
Let us start with discussing
the scalar potential for the fields $H_1$, $H_2$ and $N$, which is given by $V=V_{\rm tree}+V_{\rm 1-loop}$ as
\begin{equation}
V_{\rm tree}=V_F+V_D+V_{\rm soft}\ ,
\end{equation}
where
\begin{eqnarray}
V_F &=&|\lambda|^2 [|H_1 H_2|^2 + |N|^2 ( |H_1|^2 + |H_2|^2 )] + 
|k|^2 |N|^4 + (\lambda k^* H_1 H_2 N^{*2} + {\rm h.c.}) , \nonumber 
\\
V_D &=&{g_1^2 + g_2^2 \over 8}(|H_1|^2-|H_2|^2)^2 + 
{g_2^2 \over 2} (|H_1|^2 |H_2|^2 - |H_1 H_2|^2) \\
V_{\rm soft}&=&m_{H_1}^2 |H_1|^2
+ m_{H_2}^2 |H_2|^2 + m_N^2 |N|^2 \nonumber \\ 
& & +(\lambda A_{\lambda}H_1 H_2 N + {\rm h.c.}) + ({k A_k \over 3}N^3 + {\rm h.c.}) \ . \nonumber \end{eqnarray}

Hereafter we discuss the possibility of spontaneous $CP$ violation in the Higgs sector,
so that we take  the parameters
$h_t, h_b, \lambda, k, A_\lambda, A_k $  to be all real\cite{REI}. %
It is well known that the radiative corrections 
are important to analyze Higgs spectra
and also these corrections are essential to study spotaneous $CP$ violation in the
MSSM\cite{MAEKAWA} and the NMSSM\cite{BABU}. So the radiative corrections to the scalar potential 
at one-loop level are given by
\cite{CW}
\begin{equation}
\label{effect}
V_{\rm 1-loop} = {1 \over 64 \pi^2} {\rm Str} \: M^4 ({\rm ln} \: 
{M^2 \over Q^2}),
\end{equation}
where
Str denotes the supertrace defined as
\begin{equation}
\label{str}
{\rm Str}\: g(m^2)=\Sigma(-1)^{2J}(2J+1)g(m^2). 
\end{equation}
Here $m^2$ denotes the mass eigenvalues of a particle of spin $J$ and in the case 
of squarks $M^2$ means $4 \times 4$ $\tilde{t}_L, \tilde{t}^c_R$, $\tilde{b}_L, 
\tilde{b}^c_R$ mass squared matrices, which can be written as
\begin{equation}
\label{mass}
M_{{\tilde t},{\tilde b}}^2 =\bordermatrix{
& \tilde{t}_L &{\tilde{t}}^c_R &\tilde{b}_L &{\tilde{b}}^c_R \cr 
& m_{11}^2 & m_{12}^2 & m_{13}^2 & m_{14}^2 \cr 
& m_{12}^{*2} & m_{22}^2 & m_{23}^2 & m_{24}^2 \cr 
& m_{13}^{*2} & m_{23}^{*2} & m_{33}^2 & m_{34}^2 \cr 
& m_{14}^{*2} & m_{24}^{*2} & m_{34}^{*2} & m_{44}^2 \cr } \ ,
\end{equation}
where
\begin{eqnarray}
m_{11}^2&=&m_Q^2 + h_t^2|H_2^0|^2+ h_b^2|H_1^-|^2-{g_1^2 \over 12}(|H_1^0|^2 +|H_1^-|^2- |H_2^0|^2-|H_2^+|^2) \nonumber \\ 
&+&{g_2^2 \over 4}(|H_1^0|^2-|H_1^-|^2-|H_2^0|^2+|H_2^+|^2), \nonumber \\ 
m_{12}^2&=&h_t(A_tH_2^{0*}+\lambda NH_1^0), \nonumber \\ 
m_{13}^2&=&-h_t^2H_2^{0*}H_2^+-h_b^2H_1^{-*}H_1^0+{g_2^2 \over 2}(H_2^+H_2^{0*}+H_1^{-*}H_1^0), \nonumber\\ 
m_{14}^2&=&-h_b(\lambda N H_2^{+}-A_bH_1^{-*}), \nonumber\\ 
m_{22}^2&=&m_T^2+h_t^2(|H_2^0|^2+|H_2^+|^2)+{g_1^2 \over 3}(|H_1^0|^2 +|H_1^-|^2- |H_2^0|^2-|H_2^+|^2), \nonumber\\ 
m_{23}^2&=&h_t(\lambda N^*H_1^{-*}-A_tH_2^+), \nonumber\\ 
m_{24}^2&=&h_th_b(H_2^{0}H_1^{-*}+H_2^{+}H_1^{0*}), \nonumber\\ 
m_{33}^2&=&m_Q^2+h_b^2|H_1^0|^2+ h_t^2|H_2^+|^2-{g_1^2 \over 12}(|H_1^0|^2 +|H_1^-|^2- |H_2^0|^2-|H_2^+|^2), \nonumber \\ 
&+&{g_2^2 \over 4}\left(-|H_1^0|^2
+|H_1^-|^2+|H_2^0|^2-|H_2^+|^2\right), \nonumber \\ 
m_{34}^2&=&-h_b(A_bH_1^{0*}+\lambda NH_2^0), \nonumber \\ 
m_{44}^2&=&m_B^2+h_b^2(|H_1^0|^2+|H_1^-|^2)-{g_1^2 \over 6}(|H_1^0|^2 +|H_1^-|^2- |H_2^0|^2-|H_2^+|^2) \ . \nonumber
\end{eqnarray}
The mass parameters $m_Q,m_T,m_B$ are the soft supersymmetry breaking 
squark masses.
Here the parameters $A_t$ and $A_b$ are the soft supersymmetry breaking ones corresponding to 
the first 
two terms of the superpotential Eq.(\ref{W}); 
\begin{equation}
V_{\rm soft}=A_th_t({\tilde t}_L{\tilde t}^c_RH_2^0 -{\tilde b}_L{\tilde t}^c_RH_2^+)+ A_bh_b({\tilde t}_L {\tilde b}^c_RH_1^{-}-
{\tilde b}_L {\tilde b}^c_RH_1^{0})+{\rm h.c.}.
\end{equation}
We  also take $A_t$ and $A_b$ to be real in the present 
spontaneous $CP$ violation scenario.
\par
In order to realize spontaneous $CP$ violation in the Higgs sector, it is necessary to have nonzero complex 
VEVs for $H_1^0,H_2^0$ and $N$. We define VEVs of Higgs fields as
\begin{eqnarray}
& &\langle H_1^{0} \rangle=v_1 e^{i\theta},\quad \langle H_2^0 \rangle=v_2,\quad \langle N 
\rangle=xe^{i\xi/3}, \nonumber\\
& &\langle H_1^{-} \rangle=0, \quad \langle H_2^{+}\rangle=0, 
\end{eqnarray}
where $v_1, v_2$ and $x$ are all real and positive parameters.
\par
In our scenario, Higgs sector can be parametrized in terms of 11 free parameters:
the soft Higgs masses $m_{H_1}, m_{H_2}, m_N$, $\tan\beta$, $x$,
phases of VEVs $\theta, \xi$,
the trilinear couplings in the superpotential $\lambda$ and $k$ and the soft scalar masses
$A_\lambda$ and $A_k$.
The radiative corrections $V_{\rm 1-loop}$ due to top, stop, bottom and sbottom loops
contain the soft top mass $A_t$ and the soft bottom mass $A_b$ and the squark mass parameters $m_Q, m_B$, and $m_T$.
Then we have 16 parameters in total.
By minimizing the Higgs potential with respect to the three VEVs and two phases, we can eliminate 5 
parameters which are $m_{H_1}, m_{H_2}, m_N$, $k$ and $\xi$ by 
the equations
\begin{equation}
\label{VEV1}
{\partial \over \partial v_i}V=0 \quad (i=1,2), \qquad {\partial \over \partial x}V=0 ,
\end{equation}
and
\begin{equation}
\label{VEV2}
{\partial \over \partial \theta}V=0 , \qquad {\partial \over \partial \xi}V=0. 
\end{equation}
Then  there remains 11 parameters which determine the masses and couplings of the five neutral and 
the charged Higgs bosons.
\par
We can expand the neutral Higgs fields around their minimum points as 
\begin{eqnarray}
H_1^{0}&=&v_1e^{i\theta}+{1 \over \sqrt{2}}e^{i\theta}(S_1+i\sin\beta A) \ ,\nonumber \\
H_2^0&=&v_2+{1 \over \sqrt{2}}(S_2+i\cos\beta A) \ , \\ 
N&=&xe^{i\xi/3}+{1 \over \sqrt{2}}e^{i\xi/3}(X+i Y) \ ,\nonumber 
\end{eqnarray}
where the five components are described as \begin{eqnarray}
S_1&=&\sqrt{2}(\cos\theta{\rm Re}H_1^0-\sin\theta {\rm Im}H_1^0) \ ,\nonumber\\ 
S_2&=&\sqrt{2}{\rm Re}H_2 \ ,\nonumber\\
A &=&\sqrt{2}\{-\sin\beta(\sin\theta {\rm Re}H_1^0+\cos\theta{\rm Im}H_1^0) 
+\cos\beta{\rm Im}H_2\} \ ,\\
X &=&\sqrt{2}(\cos (\xi/3){\rm Re}N+\sin (\xi/3){\rm Im}N) \ ,\nonumber\\ 
Y &=&\sqrt{2}(-\sin 
(\xi/3){\rm Re}N+\cos (\xi/3){\rm Im}N). \nonumber 
\end{eqnarray}
If the $CP$ symmetry is conserved in the Higgs potential of the NMSSM, $\theta$ and $\xi$ should set to be zero
and the five neutral Higgs bosons are separated into three scalar bosons and two pseudoscalar bosons. 
The neutral Higgs mass matrix in the spontaneous $CP$ violation scenario 
induced by the one-loop effects is given as 
\begin{equation}
 M_H = \left(
\matrix{
{1 \over 2}{\partial^2V \over \partial S_1^2} & 
{\partial^2V \over \partial S_1\partial S_2} & 
{\partial^2V \over \partial S_1\partial A} & 
{\partial^2V \over \partial S_1\partial X} & 
{\partial^2V \over \partial S_1\partial Y} \cr
{\partial^2V \over \partial S_1\partial S_2} & 
{1 \over 2}{\partial^2V \over \partial S_2^2} & 
{\partial^2V \over \partial S_2\partial A} & 
{\partial^2V \over \partial S_2\partial X} & 
{\partial^2V \over \partial S_2\partial Y} \cr
{\partial^2V \over \partial S_1\partial A} & 
{\partial^2V \over \partial S_2\partial A} & 
{1 \over 2}{\partial^2V \over \partial A^2} & 
{\partial^2V \over \partial A\partial X} & 
{\partial^2V \over \partial A\partial Y} \cr
{\partial^2V \over \partial S_1\partial X} & 
{\partial^2V \over \partial S_2\partial X} & 
{\partial^2V \over \partial A\partial X} & 
{1 \over 2}{\partial^2V \over \partial X^2} & 
{\partial^2V \over \partial X\partial Y} \cr
{\partial^2V \over \partial S_1\partial Y} & 
{\partial^2V \over \partial S_2\partial Y} & 
{\partial^2V \over \partial A\partial Y} & 
{\partial^2V \over \partial X\partial Y} &
{1 \over 2}{\partial^2V \over \partial Y^2}\cr }
\right).
\label{Higgs}
\end{equation}
This matrix is diagonalyzed numerically and we obtain the physical Higgs 
fields $h_i \ (i=1 \sim 5)$.
As for the mass of charged Higgs boson in the NMSSM with spontaneous $CP$ violation, Babu and Barr presented the simple formula\cite{BABU}
\begin{equation}
m_{H^\pm}^2=M_W^2+(3r-1)\lambda^2v^2,
\end{equation} 
where $r\equiv A_\lambda/A_k$.
By using positivity condition of sub-determinants for squared mass matrix of neutral Higgs bosons 
and the local minimum condition for spontaneous $CP$-violation, 
they obtained the constraint 
\begin{equation}
0 \le (3r-1)\lambda^2 \le {1 \over 2}\lambda_1(\sqrt{1+\Delta}-1),
\end{equation}
where $\lambda_1 \equiv M_Z^2/v^2$ and $\Delta$ is a parameter given by the radiative effect 
at the limit of $m^2_{{\tilde t}_L}=m^2_{{\tilde t}_R}$ with 
neglecting the contribution from bottom and sbottom loop.
This constraint requires the upper limit of charged Higgs boson mass should be less than 110GeV. 
However, from the structure of squark mass matrix Eq.(\ref{mass}), 
off-diagonal elements, which do not exist in the analysis by Babu and Barr\cite{BABU},
  receive the contribution of $x$. 
The large $x$ raises the charged Higgs boson mass 
as shown in section 4  numerically. 
\section{Experimental constrains and the spontaneous $CP$-violation in the NMSSM}

 In the previous section we have obtained 
a $5 \times 5$ squared Higgs mass matrix $M_H$ in Eq.(\ref{Higgs}).
By diagonalizing this matrix the 
five eigenstates  of Higgs masses are derived and the five mass eigen states are
defined as
\begin{equation}
\label{Comp1}
\left(
\begin{array}{c}
h_1 \\
h_2 \\
h_3 \\
h_4 \\
h_5
\end{array}
\right)
=O
\left(
\begin{array}{c}
S_1 \\
S_2 \\
A \\
X \\
Y
\end{array}
\right),
\end{equation}
where the line of $l.h.s$ is the order of masses, $i.e.$ 
$m_{h_i}$ is lighter than  $m_{h_j}$ for $i<j$.
The orthogonal $5 \times 5$ matrix is defined as 
\begin{equation}
(O)_{ij} \equiv a_{ij}.
\end{equation}
The masses of these eigenstates should be positive.
This condition
means that the vacuum does not break QED in the charged Higgs sector.
The components $M_{13,23,15,25,45}$ of mass squared matrix $M_H$ are not zero 
when the $CP$ symmetry  is violated spontaneouly. 
The magnitudes of these components are proportional to $\sin\eta$ or $\sin\xi$, 
where angle $\eta$ is defined as
\begin{equation}
\eta \equiv {\rm arg}(H_1H_2N)=\theta+{\xi \over 3}. 
\end{equation}

In Ref.\cite{BABU}, Babu and Barr gave the analyses of the spontaneous $CP$ violation in the 
NMSSM by using the following experimental constraints;
\begin{description}
\item[(i)]
the condition
\begin{equation}
m_{h_1}+m_{h_2} > M_Z
\end{equation}
by the fact that Higgs bosons $h_1$ and $h_2$ have not been observed in the decay of 
$Z$\cite{PDG} and
\item[(ii)]
the lower mass limit is
\begin{equation}
m_{h_1} > (60{\rm GeV})(\alpha_1\cos\beta+\alpha_2\sin\beta)^2, 
\label{constraint}
\end{equation}
where $h_1 \simeq \alpha_1S_1+\alpha_2S_2$ by the experiment that the lightest boson
$h_1$ has not been observed in the decay  $Z \rightarrow h_1+Z^*
\rightarrow h_1 +l^+l^-$
\cite{ALEPH}.
\end{description}
\par 
However, we should carefully analyze these conditions in the case of spontaneous $CP$-violation
 in the Higgs sector. 
First we estimate  the coupling $g_{Zh_1h_2}$ and discuss  a possibility to be 
free from the experimental constraint (i) in  case of small $g_{Zh_1h_2}$ coupling
even if the sum of two lightest Higgs boson masses is lighter than $m_Z$.
The effective Hamiltonian for $Z \rightarrow h_1+h_2$ is 
\begin{equation}
H_{Zh_1h_2}={{g_{Zh_1h_2}} \over 2\cos\theta_W}Z_\mu(P_{h_1}^\mu-P_{h_2}^\mu),
\end{equation}
and
\begin{equation}
g_{Zh_1h_2}\equiv g_2(\cos\beta(a_{12}a_{23}-a_{22}a_{13})-
\sin\beta(a_{11}a_{23}-a_{21}a_{13})).
\end{equation}
In the case of $m_{h_1}+m_{h_2} < M_Z$, the decay $Z \rightarrow h_1+h_2$ is physically 
possible and the decay rate is given as
\begin{equation}
\Gamma(Z \rightarrow h_1h_2)={M_Z \over 16\pi}g^2_{Zh_1h_2}\lambda^{3 \over 2}
(1, x_{1},x_{2}),
\end{equation}
where the familiar function 
$\lambda(x,y,z) \equiv x^2+y^2+z^2-2xy-2yz-2zx$ and $x_{i} \equiv m^2_{h_i}/M^2_Z$.
  If  $B(Z \rightarrow h_1h_2)<10^{-7}$ the constraint (i) has no meanings, 
 since we take the experimental limit for rare decays of $Z$ to be $10^{-7}$\cite{PDG}.
 If the case $m_{h_1}+m_{h_2} > M_Z$     is realized, we should estimate the cross section for
the process $e^-e^+ \rightarrow "Z" \rightarrow h_1h_2$. 
By using the coupling constant $g_{Zh_1h_2}$ and the Hamiltonian $H_{Zh_1h_2}$ we obtain
\begin{equation}
\sigma(e^-e^+ \rightarrow h_1h_2)={\alpha g^2_{Zh_1h_2} \over 
24\sin^2\theta_W\cos^2\theta_W}{1 \over s} (|C_L|^2+|C_R|^2)
{\lambda^{3 \over 2}(1,y_1,y_2) \over (1-y_{Z})^2+{y_{Z}\Gamma_Z^2 \over 4s}},
\end{equation}
where $y_i=m_{h_i}^2/s$ and $y_{Z}=M_Z^2/s$.
\par
As for the constraint (ii) we can give the similar argument to the case (i). 
The coupling constant $g_{ZZh_1}$ and  $g_{ZZh_2}$ are given as
\begin{eqnarray}
g_{ZZh_1} &\equiv& {g_2 \over 2\cos\theta_W}M_Z\cos\beta(a_{11}+
                   a_{12}\tan\beta) \ ,\nonumber\\
g_{ZZh_2} &\equiv& {g_2 \over 2\cos\theta_W}M_Z\cos\beta(a_{21}+
                   a_{22}\tan\beta)\ ,
\label{ZZh}
\end{eqnarray}
respectively.
Then if $m_{h_1}$ and/or  $m_{h_2}$ are lighter than $M_Z$, the decay rate is 
\begin{eqnarray}
\Gamma(Z \rightarrow h_i l^+l^-)&=&{1 \over 96\pi^3}{g^2_{ZZh_i}g^2_{Zl^+l^-} \over M_Z}
(|C_L|^2+|C_R|^2)       \nonumber\\
& &\int_\rho^{1+\rho^2 \over 2}{1+\rho^2-2x \over (\rho^2-2x)^2+\Gamma_Z^2/(4M_Z^2)}
(x^2-\rho^2)^{1/2}dx, 
\end{eqnarray}          
where
$\rho=m_{h_i}/M_Z$, $x=E_{h_i}/M_Z$, $g_{Zl^+l^-}=2e/\sin 2\theta_W$,
$C_L=-{1 \over 2}+\sin^2\theta_W$ and $C_R=\sin^2\theta_W$.
This decay rate should be lower than the experimental upper bound $\Gamma^{\rm exp}$,
 which is equivalent to $B(Z \rightarrow h l^+l^-)<1.3\times 10^{-7}$ 
 at $m_h=60$GeV in the SM\cite{PDG}:
 \begin{equation}
 \Gamma(Z \rightarrow h_i l^+l^-) < \Gamma^{\rm exp} .
 \label{Bbound}
\end{equation}
\par
For $h_1$ and $h_2$, we use this constraint instead of Eq.(\ref{constraint}) in our spontaneous CP violation
scenario. It is noted that the constraint of Eq.(\ref{constraint}) is weaker
than ours  because it does not take into account the phase space integral.
It is found that our constraint almost rules out solutions given by Babu and  Barr\cite{BABU}.
These constraints for the masses $m_{h_i}$ are discussed  numerically in the next section.
\par
Summarizing the above arguments, we use the following experimental constraints in the next section;
\begin{description}
\item[A]
if the sum of lightest Higgs bosons $m_{h_1}$ and $m_{h_2}$ is lighter than $m_Z$, the branching ratio
$B(Z \rightarrow h_1h_2)$ should be less than $10^{-7}$ or the sum of $m_{h_1}$ and $m_{h_2}$ should
be larger than $m_Z$ and 
\item[B]
for $h_1$ and $h_2$, both of $B(Z \rightarrow h_1l^+l^-)$ and $B(Z \rightarrow h_2l^+l^-)$ should be smaller than 
$1.3 \times 10^{-7}$. 
Hereafter we call the former constraint as constraint {\bf B1} and the latter 
as constraint {\bf B2}.
\end{description}
\section{Numerical results on the spontaneous $CP$ Violation}
In this section we analyze  about the parameters $\tan\beta, \lambda, \eta, 
A_\lambda,A_t \ etc.$ in the spontaneous $CP$-violation
scenario numerically.
Assuming the perturbation remains valid up to the unification scale the couplings 
$\lambda$ and $k$ are restricted by their fixed points as pointed out by Ellis et al. 
in Ref.\cite{RADIATIVE} such as 
\begin{equation}
\label{Region}
|\lambda| \le 0.87, \quad |k| \le 0.63.
\label{k}
\end{equation}
We use these theoretical constraints to restrict the parameters in the 
followings because the spontaneous $CP$ violation gives no change 
for the renormalization group equation of the real 
parameters $\lambda$ and $k$\cite{RG}.

As mentioned in section 2, the minimization conditions Eqs.(\ref{VEV1},\ref{VEV2})
of Higgs potential determine
the soft Higgs masses $m_{H_1},m_{H_2},m_N$, the phase $\xi$ and $N^3$ coupling constant $k$.
The parameters $\xi$ and $k$ are given by 
\begin{equation}
D\sin\eta\cos\xi-(D\cos\eta+F)\sin\xi=0
\end{equation}
and
\begin{equation}
\label{Keq}
k={1 \over \lambda\sin(\eta-\xi)}\left(
{1 \over 2v_1 v_2 x^2}{\partial V_{\rm 1-loop} \over \partial\eta}-E\sin\eta \right),
\end{equation}
respectively,\
where the definition of $D, E$ and $F$ are followed by Ref.\cite{BABU} as 
\begin{equation}
D = \lambda k, \quad
E = {\lambda A_\lambda \over x}, \quad
F = {k A_k x \over 3v_1v_2}.
\end{equation}
We use quark masses and the coupling constants as 
\begin{eqnarray}
m_t&=&174{\rm GeV}, \quad m_b=4.2{\rm GeV},\quad g_1= 0.357, \nonumber \\ g_2 &=& 0.625, 
\quad h_t = 174.0/v_2, \quad h_b = 4.2/v_1. 
\end{eqnarray}
The parameters $A_t$ and $m_Q$ are given in order to 
 satisfy the necessary condition not to break color symmetry in 
the squark sector\cite{RG}\cite{RADIATIVE}. The remaining parameters $\ A_b,\ m_T,
\ m_B$ are fixed by the arguments of fixed point analyses with the assumption of GUT 
scale universality\cite{RG} as
\begin{equation}
\label{Ab}
A_b = 1.1A_t, \quad m_T = 0.95m_Q, \quad m_B = 0.98m_Q, 
\end{equation}
\noindent
where the renormalization point is taken as $Q = 3.0{\rm TeV}$. 
Under the above mentioned experimental
constraints A and B we search the relevant parameter region. 
 The allowed parameter ranges are rather narrow.
 In order to compare our result with
  the one given by Babu-Barr \cite{BABU},
 we show  the following 
  typical set of parameters, which satisfy constraints A and B, are
 \begin{eqnarray}
 \label{sol}
 \eta&=& 1.275, \quad \tan\beta = 1.0, \quad \lambda = 0.16,  \quad
  A_k = 2.9v, A_\lambda/A_k=0.8 \nonumber\\
  x &=& 3.8 v,  \quad A_T=1 {\rm TeV}, \quad m_Q=3 {\rm TeV} \ ,
 \end{eqnarray}
 \noindent
 where the parameter $k$ takes the value $-0.612$. 
In this case the Higgs masses are obtained as 
\begin{eqnarray}
m_{h_1}&=&35.8{\rm GeV}, \quad m_{h_2}=57.0{\rm GeV}, \quad m_{h_3}=177{\rm GeV}, 
\\ m_{h_4}&=&671{\rm GeV}, 
\quad m_{h_5}=785{\rm GeV}. \nonumber
 \end{eqnarray}
 \noindent
 The constraint B  is  much severer than the constraint  (ii)
  which Babu-Barr used  \cite{BABU}.
 The allowed regions obtained by Babu-Barr are almost excluded if we use constraint B.
For example, if   $\eta$ is shifted with only $\pm 0.01$, 
the solution does not satisfy the constraint B.
Then, one should shift $\lambda$ with $\pm 0.05$  in order to get allowed solution.
Thus, the allowed parameter set is very restrictive in contrast with the result given by Babu-Barr\cite{BABU}. 
We will show the results of other parameter dependence later.
\par   
For the case of parameters in Eq.(\ref{sol}), the components of each Higgs boson are given as 

\begin{equation}
\label{Comp}
\left(
\begin{array}{c}
h_1 \\
h_2 \\
h_3 \\
h_4 \\
h_5
\end{array}
\right)
=
\left(
\begin{array}{ccccc}
0.255 & 0.058 & 0.965 & 0.012 & 0.028 \\ 
0.950 & 0.167 & -0.262 & 0.002 & 0.037 \\ 
-0.177 & 0.984 & -0.013 & -0.003 & 0.031 \\ 
0.005 & 0.005 & 0.025 & 0.999 & -0.031 \\ 
-0.037 & -0.038 & -0.016 & 0.032 & 0.998
\end{array}
\right)
\left(
\begin{array}{c}
S_1 \\
S_2 \\
A \\
X \\
Y
\end{array}
\right),
\end{equation}
\noindent
 where (1,1), (1,2), (2,1) and (2,2) components are same signs.
 Since the two terms in the {\it r.h.s.} of Eq.(\ref{ZZh}) are additive,
 the coupling constants  $g_{ZZh_{1}}$ and $g_{ZZh_{2}}$ are not remarkably reduced.  
 This situation is different from that in the MSSM, where couplings are somewhat reduced.
Thus, the constraint B for the NMSSM is severer than the one for the MSSM.
 It is remarked that the lightest Higgs state mainly consists of pseudoscalar
component $A$ as shown in Eq.(\ref{Comp}).
\par
In figure 1 we give the cross section for $e^-e^+ \rightarrow h_1h_2$ from 
the threshold to $\sqrt{s}/2$=200GeV and at the energy of LEP1.5
the production cross section is about 0.8pb in the case of Eq.(\ref{Comp}). \\

\begin{center}
 \unitlength=0.7 cm
 \begin{picture}(2.5,2.5)
 \thicklines
 \put(0,0){\framebox(3,1){\bf Fig.1}}
 \end{picture}
\end{center}
\par
For the charged Higgs boson, its squared mass is given by taking the coefficient of the 
twice derivative of
$V=V_F+V_D+V_{\rm soft}+V_{\rm 1-loop}$ by $H^-$ and $H^+$ using Eqs.(3,\ref{str},\ref{mass}), 
where the physical charged Higgs is defined as
\begin{equation}
h^+\equiv \cos\beta H_2^{+}-\sin\beta H_1^{-}. 
\end{equation}
The numerical results for the parameters in Eq.(\ref{sol})  are
\begin{equation}
m_{h^\pm}=721{\rm GeV} ,
\end{equation}
\noindent which depends crucially on the squark mass $m_Q$.
We show the $m_Q$ dependence of the charged Higgs mass in figure 2,
in which other parameters are fixed as in Eq.(\ref{sol}).
The upper bound of $m_Q$ is given by $|k|<0.63$ in Eq.(\ref{k})
and lower bound by constraint B.
Thus, $m_Q$ should be larger than $3$TeV.
The predicted charged Higgs mass is too large to 
detect this boson at LEP2 and this mass becomes free from the
constraints of  $b \rightarrow s\gamma$ experiment\cite{CLEO}.
\par
\begin{center}
 \unitlength=0.7 cm
 \begin{picture}(2.5,2.5)
 \thicklines
 \put(0,0){\framebox(3,1){\bf Fig.2}}
 \end{picture}
\end{center}
\par
It is noticed that the sum of two masses is almost constant around $93$GeV 
 even if the other parameter set which fulfills the constraints A and B is taken.
 Therefore, these two Higgs bosons will be observed at LEP2 experiment 
in the near future.
In the present study we obtain rather lighter Higgs masses compared to the case without spontaneous
$CP$ violation in the NMSSM\cite{RADIATIVE}\cite{PANDITA}. 
This circumstances are  understood by the Georgi-Pais theorem for 
the radiative symmetry breaking phenomena\cite{GP}. 
It is also noted that the two lower Higgs masses are almost independent 
of the parameter x, where other parameters are fixed as in Eq.(\ref{sol}).
\par
It may be useful to comment on the value of $\tan \beta$.
There is no solution for the spontaneous $CP$ violation in the range of $\tan \beta>1$ through the numerical 
analyses.
In case of the MSSM, the arguments on electroweak symmetry breaking 
and the top Yukawa coupling lead to the allowed ranges for $\tan\beta$
as $1.0 \le \tan\beta \le 1.4$\cite{TANB} 
 although the large top quark mass does not prefer $\tan\beta\simeq 1$ 
in the RGE analyses of the Yukawa couplings. 
If $\tan\beta= 1$ is completely ruled out in SUSY, 
our scenario could not be realized
for the $CP$ violation. 
Thus, the value of $\tan\beta$ is the critical quantity for our scheme.
\par
So we investigate the available $x$ region being consistent with
the current experimental constraints A and B, where the parameter
$A_\lambda$ and $A_t$ are freely adjusted with the fixed value of $\tan\beta=1$.
It is found that the solutions exist for $x \ge 2 v$ and we show 
the typical
solution for $x=20 v$ as an example of large $x$ case for the comparison
of the relatively small  $x$ case Eq.(\ref{sol}). 

\begin{eqnarray}
 \label{sol1}
 \eta&=& 1.3, \quad \tan\beta = 1.0, \quad \lambda = 0.16,  \quad
  A_k = 16v, \quad A_\lambda=12.5 v \nonumber\\
  x &=& 20 v,  \quad A_T=1 {\rm TeV}, \quad m_Q=3 {\rm TeV} \ .
 \end{eqnarray}
 \noindent
 
In this case the Higgs masses are obtained as 
\begin{eqnarray}
m_{h_1}&=&35.9{\rm GeV}, \quad m_{h_2}=57.2{\rm GeV}, \quad m_{h_3}=177{\rm GeV}, 
\\ m_{h_4}&=&3584{\rm GeV}, 
\quad m_{h_5}=4261
{\rm GeV}. \nonumber
 \end{eqnarray}
The charged Higgs masses is  $721$GeV, which is not changed
 as far as $m_Q=3{\rm TeV}$ is fixed.
\par
The allowed region of 
 $A_t - x$ plane is shown in figure 4, in which the inside region of the triangle
is allowed.  
It is emphasized that $A_t=0$ is not allowed.
In other words, the full radiative correction at one loop level, which Babu-Barr
did not take into consideration, is significant to study spontaneous $CP$ violation in the NMSSM.
\par
\begin{center}
 \unitlength=0.7 cm
 \begin{picture}(2.5,2.5)
 \thicklines
 \put(0,0){\framebox(3,1){\bf Fig.3}}
 \end{picture}
\end{center}
\par
In Ref.\cite{BABU}, they analyzed the spontaneous $CP$ violation and obtained the
region of $\lambda$ versus $\cos\eta$. 
The available region of $\lambda$ and $\cos\eta$ is not so similar 
to our results as mentioned above.
This shows that the constraint B 
is also important  as well as  the full radiative correction at one loop level.
\par
Without spontaneous $CP$ violation the Higgs 
masses and other parameters in the NMSSM
are widely analyzed by many authors\cite{PANDITA}. It is well known that the 
NMSSM with radiative correction 
yields the heavier mass for the lightest $CP$ even scalar to be around 130GeV 
independently on the top quark mass as shown by Elliot et al. in Ref.\cite{PANDITA}.

\section{Summary and Discussion}
We have studied the spontaneous $CP$ violation 
in the NMSSM by including the 
full one-loop radiative effects into the Higgs  potential.
The parameter region being compatible with the current lower bounds 
for Higgs masses has beenanalyzed.
\par
The experimental upper bound $B(Z \rightarrow h l^+l^-)$ gives the very severe
 constraints on the solution of spontaneous $CP$ violation.
 The available region of parameters are very narrow.
   We have obtained the large spontaneous $CP$ violation as $ \eta \simeq 1.3 $. 
The solution only exists around $\tan\beta \simeq 1.0$ and 
in the vicinity of 0.16 for the coupling $\lambda$.
\par
The upper limit of the lightest neutral Higgs $h_1$ is 36GeV 
for all available parameter regions.
Also the total mass of the lightest $h_1$ and
the second lightest Higgs boson $h_2$ is almost constant 
and around 93GeV.
The charged Higgs mass is around $700$GeV, which depends on $m_Q$.
The predicted charged Higgs mass is too large to 
detect this boson at LEP2 and this mass is free from the
constraints of $b \rightarrow s\gamma$ experiment.
\par
However, if the experimental upper bound $B(Z \rightarrow h l^+l^-)$  will be improved
  in factor 1.5, one has no more solution of spontaneous $CP$ violation
   in the NMSSM.
\par  
Since $CP$ violation in the Higgs sector 
does not occur in the MSSM without a gauge singlet Higgs field $N$, 
$CP$ violation is an important signal of the existence of the gauge singlet Higgs field.
The lightest Higgs mass in the NMSSM without spontaneous $CP$ violation 
could be larger than 130GeV and
it is expected that the LEP2 experiment will give the solution on the 
possibility of spontaneous $CP$ violation in the Higgs
sector.
In the present case of the Higgs sector, 
the analyses of the electron and the neutron EDM 
and the production and the decay of Higgs state mixed with scalar 
and pseudoscalar components will 
be given in the forthcoming paper. 
\vskip 1 cm
\noindent
{\bf Acknowledgments}\par
This research is supported by the Grant-in-Aid for Scientific Research, Ministry of Education,
Science and Culture, Japan(No.06640386, No.07640413). 

\clearpage
\begin{center}
{\bf Figure Captions} \\
\end{center}
\noindent
{\bf Fig.1}\qquad
The cross section of $e^-e^+ \rightarrow h_1+h_2$ versus $\sqrt{s}/2$ in 
the case of the solution given in Eq.(\ref{sol}). \\
{\bf Fig.2}\qquad
The $m_Q$ dependence of the charged Higgs mass.\\ 
{\bf Fig.3} \qquad
The allowed region on $A_t-x$ plane constrained by a constraint {\bf B1}
(dashed line), a constraint {\bf B2}(dotted line), a constraint {\bf A}
(dash-dotted line) and a constraint $|k|<0.63$(solid line), 
where constraits are explained in section 3 of the text.

\newpage
\begin{center}
\epsfysize=10cm
\hfil\epsfbox{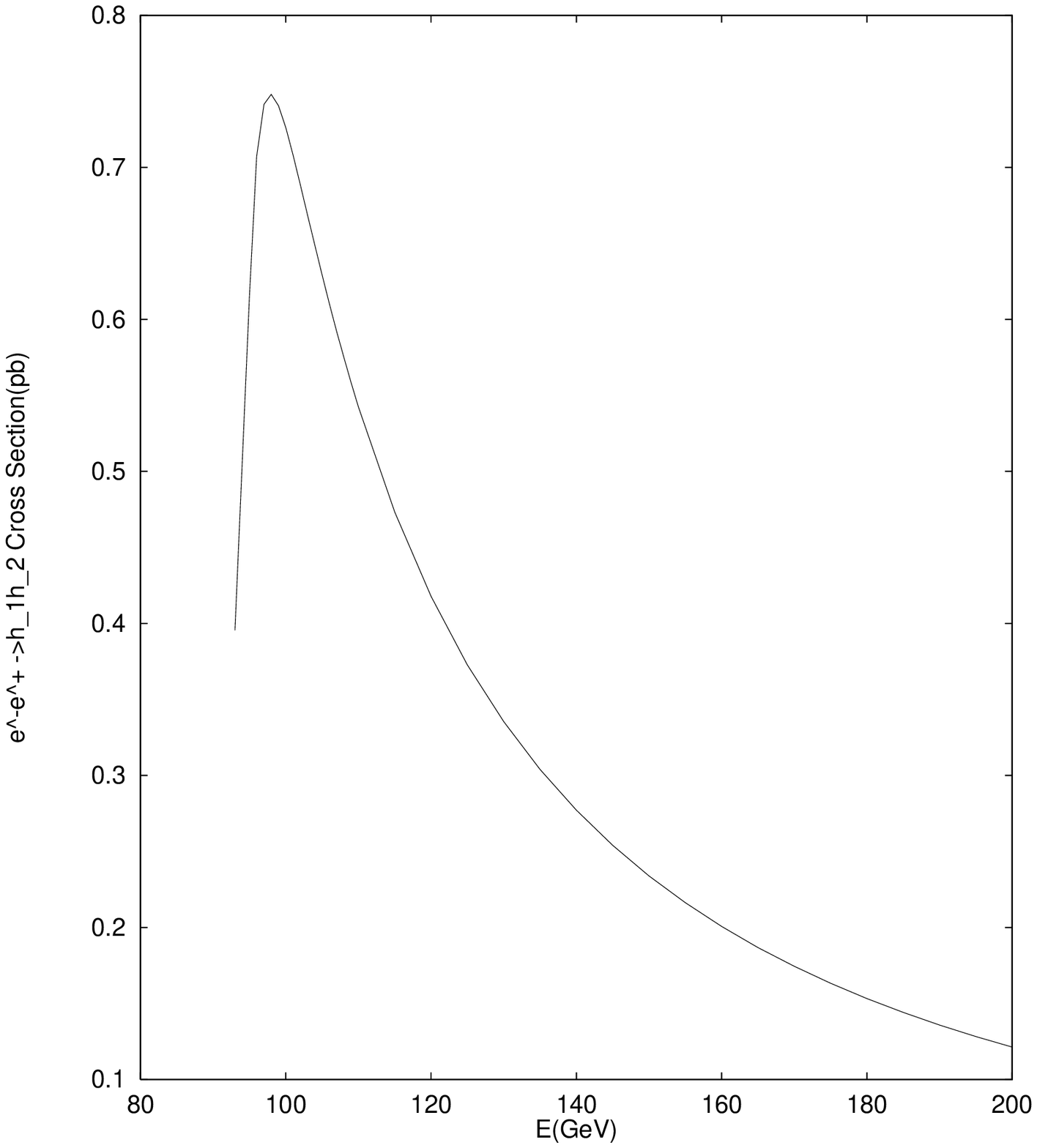}\hfill
\end{center}
\vspace{1cm}
\begin{center}
{}Fig.1  \vspace{1cm}\\
\end{center}

\begin{center}
\epsfysize=10cm
\hfil\epsfbox{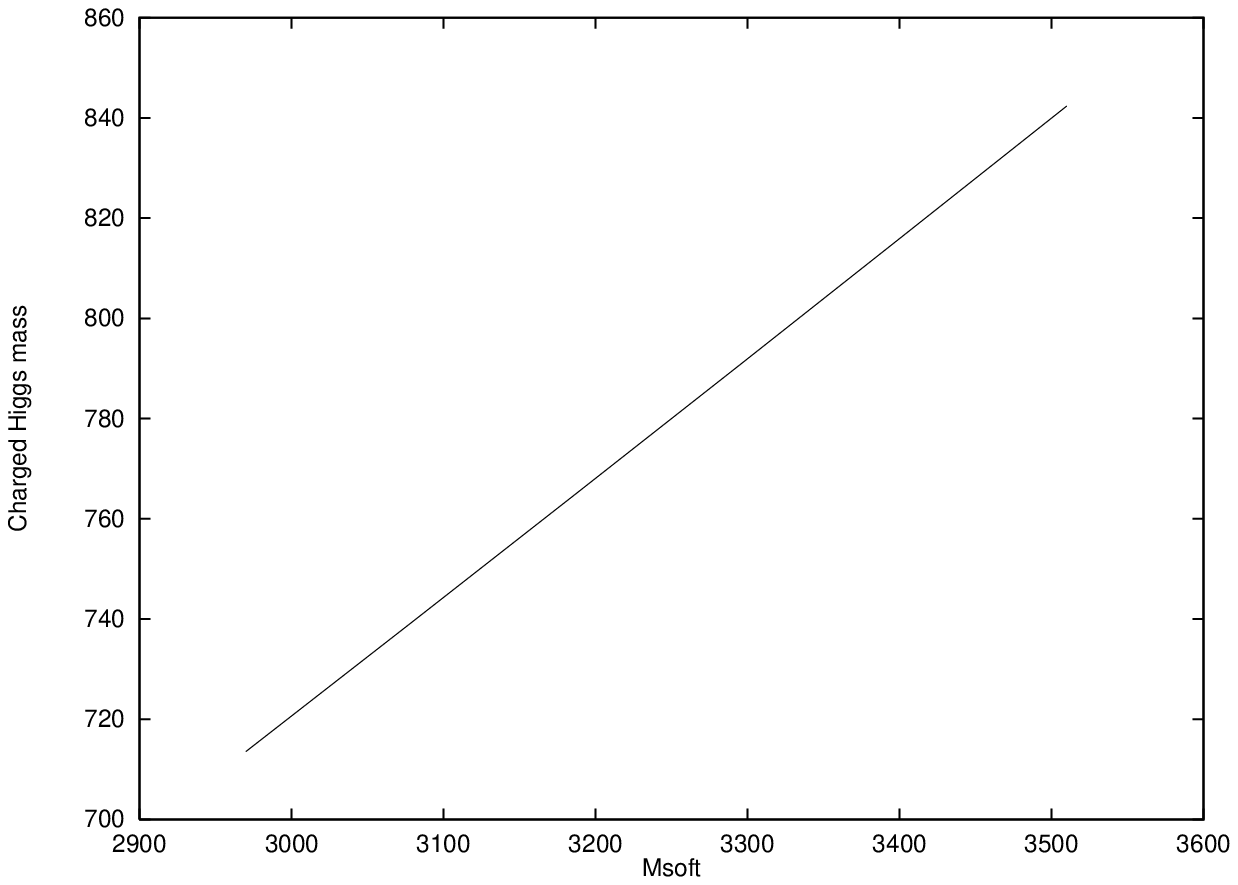}\hfill
\end{center}
\vspace{1cm}
\begin{center}
{}Fig.2
\end{center}

\vspace*{5cm}
\begin{center}
\epsfysize=10cm
\hfil\epsfbox{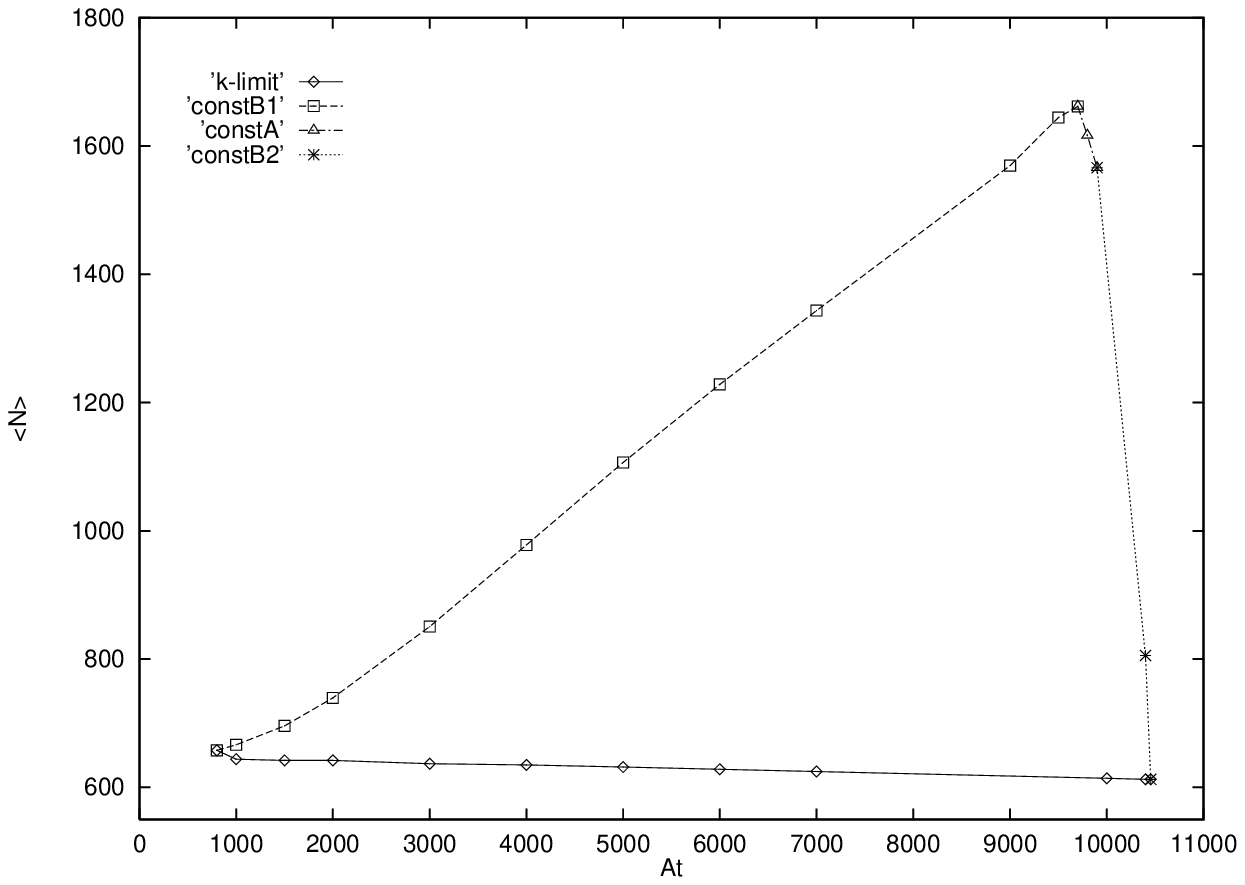}\hfill
\end{center}
\vspace{1cm}
\begin{center}
{}Fig.3
\end{center}



\begin{thebibliography}{1}
\bibitem{REI}
The possibility of explicit $CP$ violation in the Higgs sector in the NMSSM is discussed in Ref.\cite{MATSUDA}.
\end{thebibliography}
\end{document}